# Automated Phenotyping via Cell Auto Training (CAT) on the Cell DIVE Platform

Alberto Santamaria-Pang
*Artificial Intelligence*
*GE Research*
Niskayuna, USA
santamar@ge.com

Anup Sood
*Biology and Applied Physics*
*GE Research*
Niskayuna, USA
anup.sood@ge.com

Dan Meyer
*Biology and Applied Physics*
*GE Research*
Niskayuna, USA
meyer@ge.com

Aritra Chowdhury
*Artificial Intelligence*
*GE Research*
Niskayuna, USA
aritra.chowdhury@ge.com

Fiona Ginty
*Biology and Applied Physics*
GE Research
Niskayuna, USA
ginty@research.ge.com

*Abstract—* We present a method for automatic cell classification in tissue samples using an automated training set from multiplexed immunofluorescence images. The method utilizes multiple markers stained in situ on a single tissue section on a robust hyperplex immunofluorescence platform (Cell DIVE[TM], GE Healthcare) that provides multi-channel images allowing analysis at single cell/sub-cellular levels. The cell classification method consists of two steps: first, an automated training set from every image is generated using marker-to-cell staining information. This mimics how a pathologist would select samples from a very large cohort at the image level. In the second step, a probability model is inferred from the automated training set. The probabilistic model captures staining patterns in mutually exclusive cell types and builds a single probability model for the data cohort. We have evaluated the proposed approach to classify: i) immune cells in cancer and ii) brain cells in neurological degenerative diseased tissue with average accuracies above 95%.

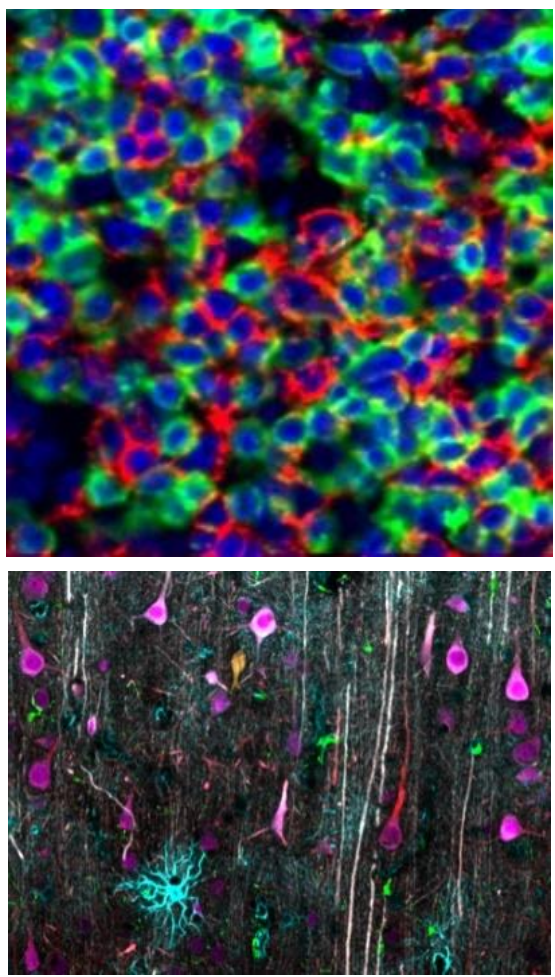

**Figure 1.** Color images for (top) immune cell markers for CD20 (red), CD3 (green), and Nuclei DAPI (blue); (bottom) brain-cell-relevant markers: NeuN (purple), non-phosphorylated neurofilament protein (SMI-32, red), GFAP (aqua), Iba1 (green), calretinin (yellow) and MAP2 (gray).

## I. Introduction

Pathologists strive for accurate and reproducible recognition of cellular patterns in tissue samples to characterize pathological diseases, with phenotype cell counting and classification as a major criteria to understand complex biological processes. However, phenotype cell classification is intrinsically difficult due to various process variables (sample preparation, non-specific staining, signal bleed through) as well as tissue structural and molecular complexity, such as dense clusters of cells and the need for multiple markers to define individual cell types. Figure 1 provides examples of overlapping marker signals from cells of different types in oncology (top) and neurology (bottom). Several multiplexed immunofluorescence methods have been described that provide the necessary multiplex staining data for cell classification. When classifying cells, manual annotation is the most common approach in selecting relevant cells that display a specific marker. Such manual annotation allows the building of recognition models to assist with cell counting and phenotyping in very large data sets.

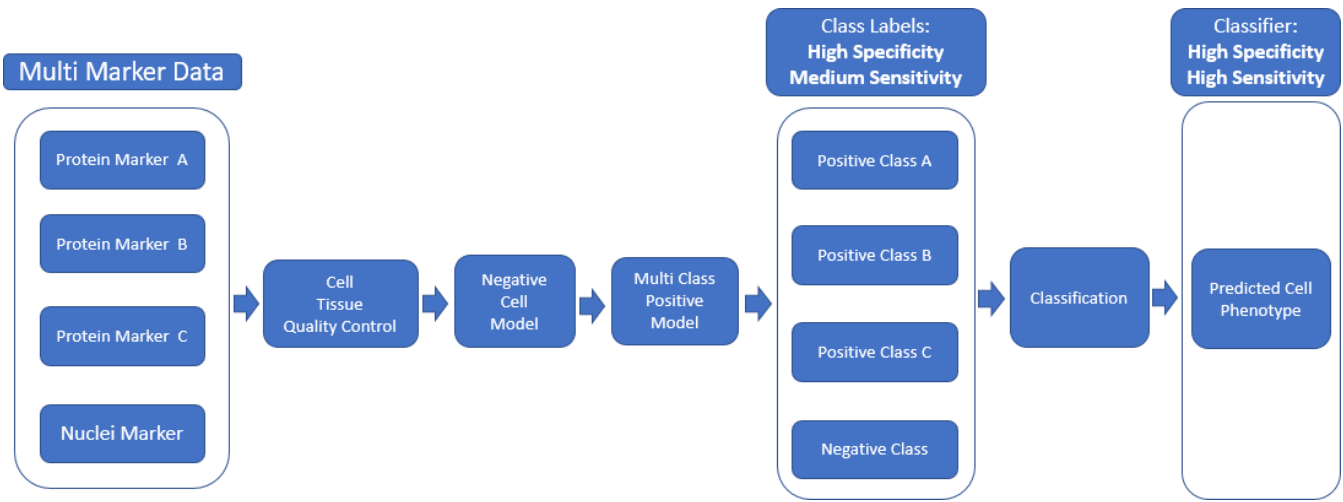

**Figure 2**. Schematic of the Cell Auto Training (CAT) algorithm to generate automated label samples of different cell types.

The most common practice for manual cell annotation relies on identifying the presence or absence of proteins via antibody-based staining: a human expert classifies and annotates strongly-stained cells in different "positive" classes (a positive class: e.g., "Marker+") from other weakly-stained cells (a negative class: e.g., "Marker–"). The "negative" class needs to be assigned to cells that have either low marker intensity or have minimum overlap with the compartment of interest. Nevertheless, the annotation task is not only laborious and time-consuming, but also may be potentially biased, making it extremely challenging to implement across thousands of regions of interest. Substituting this manual task with an automated and objective cell annotation algorithm would significantly improve throughput, accuracy and reproducibility.

## II. Literature Review

Automated object classification is one of the main research areas in machine intelligence. Despite the extraordinary progress that has been made over the last twenty years, it remains an unmet need [1]. Depending on the type of application, two major computational methods have been the most successful: i) unsupervised data driven methods that try to create an inference model which captures the underlying probability distribution from the data [2], typically they are posed as clustering or association problems; ii) supervised methods that approximate a mapping function from input and output variables of the form: $y=f(x)$, where selection of the mapping function $f$ is expressed as a classification or regression problem [3]. Both approaches offer advantages depending on the data and end application. Unsupervised methods do not require training data and therefore can be applied in very large datasets without the need for data annotation. Supervised methods have demonstrated exceptional performance in medical and biomedical applications, overperforming their unsupervised and data driven counterparts [4]. While they provide practical solutions in small datasets, scaling fully supervised algorithms to very large and heterogenous datasets is extremely difficult due to the large amount of annotated training data that human experts must provide [5].

There are several methods to address automatic labeling for recognition and classification in very large datasets including image, text and speech. A two-step approach for database record-pair classification is presented in Christen [6]. The first step automatically produces training data and then it is used in a second step to train a binary classifier. In Judea [7] a method for unsupervised training set generation of labeled training data for automatic terminology acquisition is presented. The approach includes a candidate classifier and a conditional random field and is demonstrated in technological knowledge sources for patents. Schoeler [8] presented the TRANSCLEAN algorithm for natural image classification using multiple languages and context-check to generate an image dataset to further use in object category modeling for classification. A similar application to handwritten character recognition was presented in Vajda [9] where semi-automated ground truth generation was used in large vocabulary continuous speech recognition (reported in [10]). The semi-automated ground truth generation of unlabeled data was recognized using a seed model and the hypotheses from the recognition system were used as transcriptions for training. The performance of unsupervised and directed manual transcription training was evaluated from Mandarin transcription task showing promising results.

In the context of biomedical imaging, multiple methods have been used for automated cell classification. In Xu [11] a weakly supervised algorithm using multiple clustered instance learning for i) image-level classification (cancer vs. non-cancer image), ii) medical image segmentation (cancer vs. non-cancer tissue), and iii) patch-level clustering (different classes) is presented using colon cancer cytology images. Wiliem [12] presented a method for automated classification of human epithelial cell types using indirect immunofluorescence via the cell pyramid matching algorithm using Kernel-based methods. Similarly, Nazlibilek [13] used automatic segmentation, counting, size determination and classification for white blood cells from histopathology images using principal component analysis and a nearest neighbor classifier.

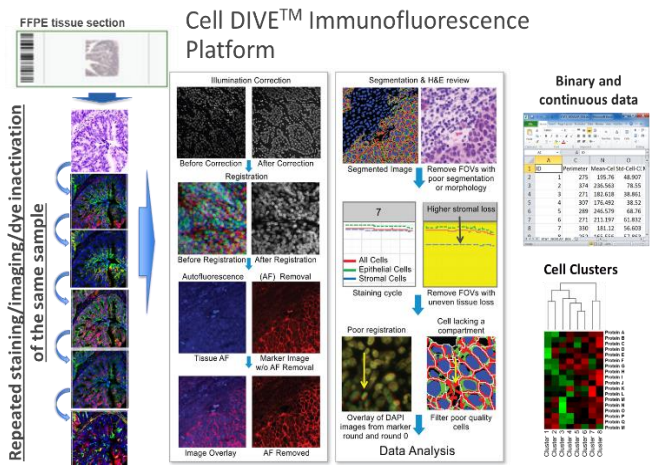

**Figure 3:** An overview of Cell DIVE™ platform showing various steps of image generation processing, data quality control and data generation.

The previous methods address automated or semi-automated classification for object, text or speech recognition tasks. However, to the best of our knowledge, no methods have been reported to specifically mimic the intrinsic mechanism with which human experts define multiple cell phenotypes and training sets from very large datasets with high levels of data variability.

The main contribution of this paper is the introduction of a new method for automatic generation of high quality training data for cell classification. The method mimics how human experts select the most relevant cell phenotypes according to marker intensity and cell morphology. Then, using this high quality training dataset, a machine learning algorithm is used to generate a probabilistic model and classify cell phenotypes. We refer to this method as Cell Auto Training (CAT) and we demonstrate its efficacy in multiplexed immunofluorescence images. The rest of this paper is organized as follows: Section III describes our proposed methods. Results and conclusions are presented in Sections IV and V respectively.

### III. METHODS

The proposed algorithm automatically generates a multi-class training set for different cell phenotypes. This training set contains cell samples with high specificity and medium sensitivity. To achieve this, several complementary sources of biology driven information are used, such as: i) marker intensity, ii) marker-cell colocalization, iii) cell morphology and iv) cell-data tissue quality. Then, using the automatically generated training set, a supervised classification algorithm is trained to classify all cells in the dataset (Figure 2).

#### A. Marker Staining, Imaging and Image Processing

We use a robust hyperplex immunofluorescence microscopy platform (Cell DIVE™) that allows subcellular imaging of over 60 markers in a single 5um formalin-fixed, paraffin-embedded (FFPE) tissue section [18]. The method involves iterative rounds of staining, imaging and signal inactivation followed by multiple steps of image processing including illumination correction, registration and autofluorescence removal [16]. The commercially available system allows rapid whole slide imaging and can be configured for automated 24/7 operation. Over 400 antibodies/targets including compartment-specific segmentation markers, various cell type markers (including immune and neuronal cells), pathway and structural markers have been validated to perform well on this platform using a multi-step process previously described [18]. An analytical image analysis pipeline has been optimized on this platform, providing the following capabilities: image processing, image segmentation, cell classification and image-tissue quality quantification. Images generated on this platform can be processed using a wide variety of available image analysis software tools (Figure 3).

#### B. Nuclei and Marker Segmentation and Tissue QC

For this study we used nuclear segmentation using DAPI (a stain specific for nucleic acids) signal and overlay the various marker signals (that define different cell classes) into the segmented cell nuclei to generate annotations of high specificity. To that end, we use a wavelet-based segmentation algorithm to segment cell nuclei and markers based on morphology. The algorithm uses wavelet coefficients to enhance blob-like objects and obtain a segmentation mask per object (for more details see Padfield [14]). For cell nuclei segmentation, we use the first round of staining since this state is the most preserved.

In general, a small fraction of tissue may not be fully preserved during the multiplex staining and imaging process due to tissue deformation, folding or actual tissue loss. To avoid any potential bias, we detect well preserved tissue across all staining rounds. To achieve this, we determine correlation-based metrics [15] that measure the tissue alignment of the same cell objects between successive rounds of staining, and we filter cells where tissue damage may have occurred using a correlation value threshold. Once each cell nucleus is segmented and filtered by the correlation-based quality control metrics, a unique identifier per image is assigned.

#### C. Cell Auto Training (CAT)

In different biological applications, detection of multiple and mutually exclusive cell types frequently occurs. For instance, when classifying different brain cells, one is interested in identifying: neurons, astrocytes, microglia, oligodendrocytes and other glial cells within a Region-Of-Interest (ROI). Or when performing immune cell analysis, one needs to detect various immune cell types including: T, B and various myeloid-derived cell populations, and within T cells a different number of mutually exclusive cell populations (e.g. helper T cell, cytotoxic T cell, regulatory T cell and so on), many of which require multiple markers for identification. This can be an extremely challenging and time-consuming task for human experts. Figure 4 shows a schematic of the main steps to automatically select multiple cell types.

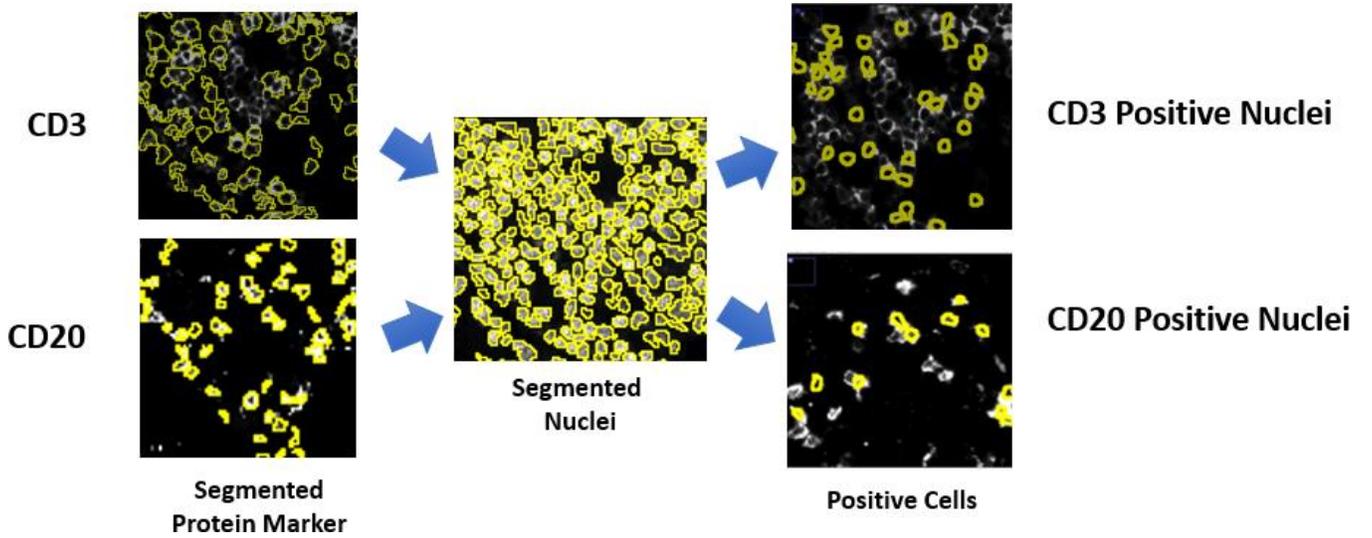

**Figure 4:** Multi-class cell selection using CD3 and CD20 protein markers in a selected ROI.

The proposed method computationally mimics how a human expert selects specific cell phenotypes using marker expression and cell morphology. Hence a subset of cells that represent the underlying cell class distribution are automatically determined. Biological hypotheses related to mutual exclusivity and co-expression for different cell targets are inferred in very large sets of multi-channel images. A multiplexed image cohort $L$ is a set of multi-channel images $L = \{L_1, …, L_n\}$ and each multi-channel image $L_j$ is a sequence of single channel images $\{I_j, M_{j,1}, …, M_{j,k}\}$ which comprises a cell nuclei image $I_j$ and a marker image $M_{j,k}$, where $1 \leq j \leq n$ and $k$ is the total number of markers.

For simplicity we denote the segmented cell nuclei $x_j^i$ as an individual cell or cell in the image $I_j$, where $1 \leq i \leq h$, and $h$ is the total number of cells. Note that cell population per image is variable according to the selected image. Let $N$ be the set of cells, we aim to estimate a function $C$ that assigns a class label to a subset of cells based on multiplexed images $L$ defined as the mapping $C: N' \to Y$, where $N'$ is a subset of the whole cell population and $Y$ is a cell class label. Automated cell labeling uses intensity and morphology to assign positive and negative cell types.

*1) Negative Cell Class Label Generation*

We approximate the overall pixel marker intensity probability distribution as the mixture of Gaussian distributions for the image $j$ and marker $k$:

$$F_{j,k}(\theta) = aP_{j,k}^F(\mu_{j,k}^F, \sigma_{j,k}^F) + bP_{j,k}^B(\mu_{j,k}^B, \sigma_{j,k}^B)$$

where $P_{j,k}^F$, $P_{j,k}^B$ are Gaussians probability distributions for foreground and background with parameters $(\mu_{j,k}^F, \sigma_{j,k}^F)$ and $(\mu_{j,k}^B, \sigma_{j,k}^B)$ respectively. Then a background probability value is assigned to each cell nucleus $x_j^i$ and image marker $M_{j,k}$ as:

$$P_{B\,i,j,k} = \frac{\sum_{(x,y)\,\in\,x_j^i} P_{j,k}^B\,(M_{j,k}(x,y) \mid \mu_{j,k}^B, \sigma_{j,k}^B)}{|x_j^i|}$$

where $|x_j^i|$ denotes cell nuclei area and $(x,y)$ denotes pixels corresponding to the cell nucleus $x_j^i$. In the previous equation, the numerator is the sum of background probabilities within the cell nucleus. The denominator is the number of pixels within the cell nucleus $x_j^i$. The previous score $P_{B\,i,j,k}$ can be interpreted as the normalized background probability of each nucleus with respect to each marker $k$. Then, negative label outputs are expressed as: $Negative\ Label\ if\ P_{B\,i,j,k} < T_k\ \forall k$, where $T_k$ is a probability threshold value for cell nuclei $i$ in image $j$ and marker $k$. Any cell nuclear object that does not meet the prior condition is then used as a candidate for positive label marker.

*2) Positive Cell Class Label Generation*

Cells that are not labeled negative are selected as candidates for positive marker labeling. A score per cell is generated assuming a portion of the signal in cell marker compartment (e.g. membrane) that overlaps with the cell nucleus. Cell membrane-surface segmentation is obtained from a variation of the wavelet-based segmentation algorithm applied to segment membrane-like objects, denoted as $m_{j,k}^i$ with respect to the marker image $M_{j,k}$. This indicates that if a cell membrane is segmented, then the likelihood for marker positive labeling increases and is defined as follows:

$$P_{F\,i,j,k} = \frac{||x_j^i| - |m_{j,k}^r||}{|x_j^i|}$$

for some membrane-like $m_{j,k}^r$ object (from marker $k$) that overlaps with the nuclei $x_j^i$. Then $P_{F\,i,j,k}$ is a probability score which measures marker positivity.

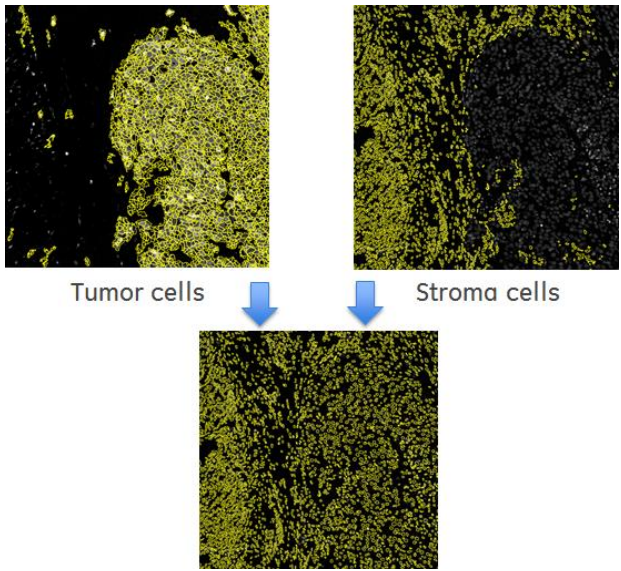

**Figure 5.** Cell nuclei segmentation. Top: tumor cells nuclei and stroma cell nuclei segmentation. Bottom: combined nuclei across the selected field of view.

We assign multiple positive cell class labels by maximizing the marker-cell expression likelihood at any given class:

$$P_k(x_j^i) = \max_{1 \leq k \leq z} P_{F\,i,j,k}$$

Then the positive cell class is obtained as:

$$C(x_j^i) = \begin{cases} k & if\ P_k(x_j^i) \geq T_k \\ Negative & if\ P_i(x_j^i) < T_k \end{cases}$$

where $T_z$, $0 \leq T_z \leq 1$, is a probability threshold value with respect to the marker $z$, $1 \leq k \leq z$. Identifying the optimal cell-marker expression across the markers provides a data sample of positive cell classes across z images. This dataset is further used to train single cell-based classifiers.

### D. Cell Type Probability Estimation

There are different methods to build a classification model from a training set and most of these techniques require a (nearly) balanced dataset. However, given that the cell phenotypes of interest are expressed in different percentages, ranging from <1% (oncology) to 30% (neuro), the negative cell class is considerably higher than the positive, leading to highly un-balanced datasets. To account for unbalanced classes and obtain a good approximation for the distribution of the positive and negative classes, we randomly sample all the negative cells with an equal number to the cells in the individual positive classes.

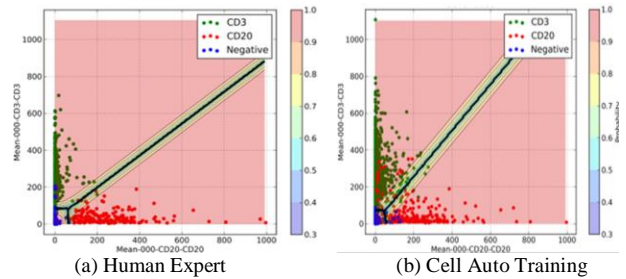

(a) Human Expert  (b) Cell Auto Training

**Figure 6.** Comparison of training sets using two markers: CD20 and CD3. Green and red points represent cells that are positive for CD3 and CD20 respectively. Blue points represent cells that do not express CD3 or C20. Figure 6(a) shows a training set, estimated classification hyperplane overlaid with estimated probabilities from a human expert, similarly, Figure 6(b) shows the result from our algorithm.

## IV. RESULTS

We have validated our automated cell classification algorithm primarily using samples derived from human colon cancer and also demonstrated preliminary application to neurological tissue from rat brain.

### A. Colon Cancer Tissue

We have applied the proposed method to the classification of immune cells in colorectal cancer. The colon cohort in this analysis was collected from the Clearview Cancer Institute of Huntsville Alabama from 1993 until 2002, with 747 patient tumor samples collected as formalin-fixed paraffin-embedded specimens. A full description of materials and methods was described recently in [15]. We restrict our analysis to 275 patient samples that correspond to grades 1 and 2; among the 747 samples and classification of B (CD20) and T (CD3) cells.

Immune cell phenotyping involves classifying stroma and epithelial tissue and single cell nuclei. We use epithelial-based markers to classify tissue areas as stroma tissue and epithelial tissue. We segment individual cell nuclei according to [14], whereas for epithelial cell nuclei, we apply our hierarchical cell and sub-cellular segmentation algorithm according to [16], [17], [18]. We then combine epithelial cell nuclei and stroma cell nuclei into a single nuclei segmentation mask. Figure 5, shows the process of combining epithelial and stroma nuclei. Once we have a single nuclei cell mask, we proceed to generate an automated training set which contains three classes: CD20, CD3 and double negative as described in Section III. Then using the automated training set, a classifier is trained according to [19].

Figure 6 shows a comparison of the training set from a human expert (Figure 6(a)) and our proposed method (Figure 6(b)). Green and red points represent cells that are positive for CD3 and CD20 respectively. Blue points represent cells that do not express CD3 or CD20. The classes can be both down-sampled and up-sampled (using synthetic decomposition methods such as SMOTE), resulting in equal populations.

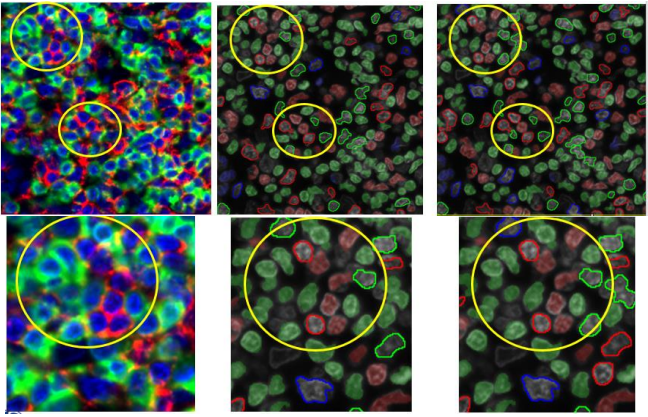

**Figure 7**. Comparison of manual vs. automated model predictions. (top) Selected ROI and (bottom) detail from ROI. Red, green and blue are CD20+, CD3+ and dual negative respectively. Cells with colored borders represent probability less of 0.9 while cells without borders represent predictions greater than 0.9.

*Validation:* Table 1 shows the estimated sensitivity and specificity of CD3 and CD20 cell classification from colorectal cancer from ~6K annotated cells. Here, we compared the model predictions with respect to the annotated cells from a human expert.

TABLE I. CELL ACCURACY IN COLON CANCER

|      | Sensitivity | Specificity | Overall Accuracy |
|------|-------------|-------------|------------------|
| CD3  | 0.98503     | 0.911599    | 0.96589          |
| CD20 | 0.897704    | 0.991111    |                  |

The overall accuracy was 96.5% for CD3 and CD20 positive cells respectively. Marker sensitivity was 98.5% (CD3) and 89.8% (CD20). Marker specificity was 91.1% (CD3) and 99.1% (CD20) respectively. Figure 7 presents a comparison of predictions derived from the manual and the cell auto training algorithm.

### B. Brain Tissue

To demonstrate applicability in brain tissue, we also applied our automated cell classification algorithm in rat brain tissue to classify neurons and microglia brain cells using NeuN and Iba1 marker respectively. Images were acquired at 20X magnification yielding a pixel size of 0.37 μm in the x-y axes. One 5 μm thick formalin-fixed, paraffin embedded (FFPE) tissue section for each case was subjected to the multiplexed staining protocol, as described above.

For brain tissue, we directly segmented cell nuclei using DAPI marker and a segmentation algorithm with parameters used in colon cancer [14]. Regarding data quality control, we retained cells that remain unshifted across all the staining rounds using correlation metrics [15].

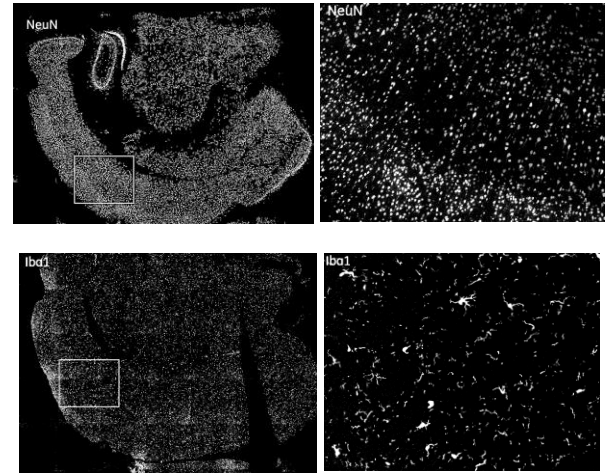

**Figure 8.** Rat brain tissue for Neuron (NeuN) and Microglia (Iba1), top and bottom respectively. Right images show details of interest from NeuN and Iba1.

Figure 8 shows an example of typical tissue staining patterns for neurons and microglia from NeuN and Iba1 markers respectively. To generate the automated training set we segmented NeuN and Iba1 as follows: given that NeuN primarily stains the nucleus and does spill out into the cytoplasm, we applied the nuclei segmentation [14] with different parameters to account for bigger blob-like objects. To achieve Iba1 marker segmentation, a variation of the method described in [20] was used. Once marker segmentation is obtained, automated cell class selection is generated for NeuN and Iba1 and negative classes respectively. Figure 9 shows a color overlay of segmented nuclei (yellow outline) overlaid with the color composite image NeuN (neurons), Iba1 (microglia) and DAPI (cell nuclei). Then we further applied the cell classification framework described in [19] to classify neurons, microglia and double negative cell classes.

TABLE II. CELL ACCURACY IN BRAIN

|           | Sensitivity | Specificity | Overall Accuracy |
|-----------|-------------|-------------|------------------|
| Neuron    | 0.963061    | 0.901235    | 0.956929         |
| Microglia | 0.962963    | 0.956608    |                  |

*Validation:* The algorithm was validated by a human expert using a random sample of 1,000 cells that is proportionately distributed according to the predicted cell classes: neurons, microglia and double negative. Our method achieved an overall classification accuracy of 95.6%. The overall sensitivity for both cell types was 96%. Specificity for neurons and microglia was 90.1% and 95.6% respectively. While we observed similar sensitivity in both neurons and microglia, we observed approximately 5% difference in specificity for both cell types.

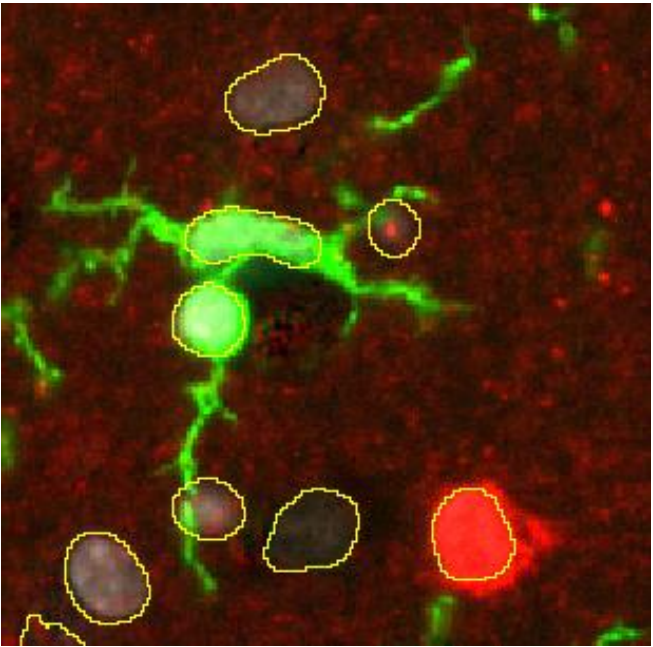

**Figure 9.** Color composite image: NeuN (neurons), Iba1 (microglia) and DAPI (cell nuclei), in red, green, grayscale respectively. Segmented nuclei overlay in yellow contours.

## V. CONCLUSIONS

We have presented a computational framework to automatically generate high quality training data for cell phenotypes. The presented framework provides the ability to automatically annotate mutually exclusive cell classes instead of using manual annotations. We demonstrated direct applicability in classifying immune cells from colon cancer and brain cells with high accuracy. The method is general since it can be used for training different classification algorithms such as classical machine learning and deep learning methods. Our approach can save significant time and cost without loss of quality.